\documentclass[a4paper,10pt]{article}

\usepackage{fullpage}
\usepackage{hyperref}
\usepackage{authblk}

\usepackage{listings}
\usepackage{xspace}
\newcommand{\MMA}{\emph{Mathematica}\xspace}
\newcommand{\eg}{\emph{e.g.}\xspace}
\newcommand{\ie}{\emph{i.e.}\xspace}

\newcommand{\mma}[1]{\lstinline{#1}}
\newcommand{\pkg}[1]{\textsc{#1}}

\newcommand{\secref}[1]{Section~\ref{#1}}

\begin{document}
	
\title{Symbolic quantum programming for supporting applications of quantum computing technologies}

\author{Jarosław Adam Miszczak}
\affil{%
	{Institute of Theoretical and Applied Informatics, Polish Academy of Sciences},
	{Bałtycka~5}, {44-100}
	{Gliwice}, 	{Poland}
	
}

%

\maketitle

\begin{abstract}
The goal of this paper is to deliver the overview of the current state of the art, to provide experience report on developing quantum software tools, and to outline the perspective for developing quantum programming tools supporting symbolic programming for the needs of quantum computing technologies. The main focus of this paper is on quantum computing technologies, as they can in the most direct way benefit from developing tools enabling the symbolic manipulation of quantum circuits and providing software tools for creating, optimizing, and testing quantum programs. We deliver a short survey of the most popular approaches in the field of quantum software development and we aim at pointing their strengths and weaknesses. This helps to formulate a list of desirable characteristics which should be included in quantum computing frameworks. Next, we describe a software architecture and its preliminary implementation supporting the development of quantum programs using symbolic approach, encouraging the functional programming paradigm, and, at the same, time enabling the integration with high-performance and cloud computing. The described software consists of several packages developed to address different needs, but nevertheless sharing common design concepts. We also outline how the presented approach could be used in tasks in quantum software engineering, namely quantum software testing and quantum circuit construction.\\[6pt]
\end{abstract}

%
\maketitle

\section{Introduction}

Quantum technologies are on the rise~\cite{harrow2017quantum}. Quantum computing promises to deliver a significant computational speed-up for many crucial problems, including optimization problems important in logistics, medicine, chemistry, and physics. This promise has been ignited by the demonstration of quantum supremacy, which confirmed that quantum computers are indeed hard to simulate even for the most powerful existing super-computers. For this reason, it is not surprising that many public \cite{qtflagship,pathfinder} and commercial \cite{ibmqroadmap,awsquantum} initiatives have been launched to support the development of quantum computing hardware and software. They result in a rapid and systematic progress in the available quantum computational resources~\cite{ibm433}. Additionally, the need for simulating quantum computing, inevitably linked to the need to develop quantum algorithms, increased interest in this technology among the companies traditionally on the forefront of high-performance computing~\cite{qhipster-intel,cuquantum-nvidia}.

This rapid progress of hardware technologies triggered a significant interest in providing software ecosystems supporting the development of application based on the technologies harnessing the potential of quantum computing. As many aspects of quantum computing require specialized software and skills, it is only natural to ask to what degree can software engineering support the manipulation of logical and computational structures required to harness the potential of quantum computing \cite{miszczak2012high,qosf-awesome-quantum-software}. Thus, the clear requirements for developing the field of quantum software engineering are  needed~\cite{ali2022when}. Additionally, quantum programming and quantum software development require specialized knowledge to be disseminated among software enthusiasts and developers, and for this purpose, quantum computing education~\cite{salehi2022computer} efforts need to be supported by the software community.

This paper is organized in the following manner. First, in \secref{sec:stat-of-art} we review the recent progress towards providing software tools supporting the development of quantum programs. We also offer some examples of software packages targeting the typical needs of the quantum software users. Next, in \secref{sec:towards} we provide an overview of the architecture supporting the symbolic development of quantum programs and describe a preliminary implementation of some elements of this architecture. To support the rationale behind developing such tools, in \secref{sec:applications} we provide a brief discussion of the problems in quantum computing technologies which could potentially benefit from the symbolic approach. Finally, in \secref{sec:conclusions} we summarize the contribution of this paper and provide some concluding remarks.

\section{Structure of the quantum software ecosystem}
\label{sec:stat-of-art}

Let us start by providing a short review of the current state-of-art in the field of quantum software. One should note that this section does not aim to provide a comprehensive listing of software packages and tools. Such lists are hosted by various organizations, including the ones that can be found at \cite{qosf-awesome-quantum-software} and \cite{quatiki-list-qc-simulators}. Additionally, recent surveys of the progress and trends in the field can be found in \cite{zhao2020quantum} and \cite{gill2021quantum}. Moreover,  an extensive comparison of quantum computing software platforms is provided in~\cite{larose2019overview}.

On the other hand, we aim to provide an overview of the approaches used for developing software supporting various aspects of quantum computer science, ranging from the basic research in quantum physics, through structures studied in quantum computing and quantum information, and closing with the specialized software development frameworks, domain-specific languages, and zero-quantum code solutions. We aim to provide the examples in all of these categories of software frameworks, and using them point out strong elements and drawbacks of the different approaches for creating quantum code development frameworks. 

\subsection{High-performance simulators}

The most straightforward approach for developing quantum software is to utilize specialized software focusing on simulating quantum physical phenomena. Among the recent additions to this family of tools supporting quantum programming, one can include several interesting items.

One of them is \pkg{Qulacs}~\cite{suzuki2021qulacs}, which is a \pkg{Python}/\pkg{C++} library for fast simulation of large, noisy, or parametric quantum circuits. As in numerous instances for this type of software, \pkg{Qulacs} was developed with efficiency in mind. To this end, it provides the efficient methods, implemented in the form of \pkg{C++} library, targeting small quantum circuits, which are simulated many times, for example to deliver the expected value of the objective function in variational quantum algorithms. Similar approach to deliver high-performance simulation engine is provided by \pkg{TornadoQSim} \cite{stratikopoulos2022tornadoqsim} developed in \pkg{Java}. In this case, the quantum simulator back-ends are accelerated on heterogeneous hardware using \pkg{TornadoVM} open source virtual machine. \pkg{TornadoVM} automatically accelerates Java programs on multicore CPUs, GPUs, and FPGAs, enabling the utilization of heterogeneous computing. 

In contrast to the two solutions mentioned above, an approach based on well-established programming languages, \pkg{Yao.jl}~\cite{luo2020yao.jl} aims to leverage the capabilities of \pkg{Julia} programming language. \pkg{Yao.jl} goal is to deliver an extensible, efficient open-source framework for quantum algorithm design. As the package is written in Julia, it tends to inherit all strengths and weaknesses of this programming language. From the perspective of this work, it is important to mention that \pkg{Yao.jl}, via \pkg{YaoSym} component, supports  symbolic computation. This is a rare example of such support among the existing quantum software frameworks. 

\subsection{Research oriented software}

Another example of quantum software package supporting symbolic manipulation is \pkg{Tequila}~\cite{kottmann2021tequila}, which provides an abstraction framework for (variational) quantum algorithms. Both in the case of \pkg{Yao.jl} and \pkg{Tequila} the support for symbolic manipulation was introduced to support automatic differentiation of objective functions used to build quantum circuits in variational quantum algorithms. However, in both cases, the support is limited by the external libraries providing symbolic capabilities. The most interesting feature of \pkg{Tequila} is that it can be used with several backbends, including \pkg{Qulacks}, mentioned above, and \pkg{Qiskit} and \pkg{Cirq}, supported by large software companies. This approach, used in \eg \pkg{ProjectQ}~\cite{steiger2018projectq,haener2018software}, ensures that the software written using the particular framework is hardware-agnostic and can be run using various quantum computing architectures. 

One of the motivations for developing quantum software supporting quantum elements is to provide support for basic research in the theory of quantum computing and quantum information. Hence, a considerable number of packages has been developed to support common elements, constructions, and protocols in these fields. Usually, developers of this kind of software packages are mostly concerned with states, unitary gates, channels, and measurements. 

Among the most popular packages to support research in quantum computing and quantum physics one can include \pkg{QuTip}~\cite{johansson2013qutip}. This package delivers a \pkg{Python} library for simulating quantum optics and quantum information.

One of the recent examples is \pkg{toqito}~\cite{russo2021toqito}, which is provided as a \pkg{Python} library for studying various objects in quantum information. Besides supporting the standard library of functions for manipulating quantum states and operations, \pkg{toqito} incorporates an optimization module to solve semidefinite programs and supporting the study of non-local games.

\subsubsection{Specialized models of quantum computing}
A significant number of quantum software packages dedicated to research is focused on a narrow area of quantum computing. Among such particular areas, the field of quantum walks attracted a considerable attention. This is in part motivated by the fact that this paradigm of quantum computing was successfully applied in the development of new efficient quantum algorithms and defining the capabilities of quantum computing.

The basic packages dedicated to supporting quantum walks support the simulations of quantum walks for 1D and 2D lattices~\cite{marquezino2008qwalk}. However, the quantum walks paradigm offered a plethora of generalizations. The recent interest in quantum walks is mainly focused on the ability to express the general form of quantum and classical evolution on graphs. This model has been used to create a package \pkg{QSWalk.jl}~\cite{glos2019qswalk.jl} which provides a high-performance analysis of quantum stochastic walks for Julia programming language. The functionality of the package is very similar to the functionality of package \pkg{QSWalk.m}~\cite{falloon2017qswalk}.
Compared with \pkg{QSWalk.m}, \pkg{QSWalk.jl} has two main advantages. From the usability perspective, it can be used to describe quantum stochastic walks in the local, as well as global regime. More importantly, by leveraging the capabilities of \pkg{Julia} language, it enables the user to seamlessly utilize parallel computing capabilities. This second aspect is of importance for any quantum software framework.

\subsubsection{Simulation of physical phenomena}

Another category of software packages important from the perspective of quantum software development is focused on the simulation of physical phenomena underlying quantum hardware. One should keep in mind that this type of software is different from the general-purpose simulators, such as \pkg{Qulacs} or \pkg{Yao.jl}, where one is mostly interested in the simulation of the logical structure of quantum programmes. On the contrary, the specialized software of this category focuses on selected physical processes, thanks to which it allows  the exact creation of the physics of specific phenomena.

A prominent example of this type of software is given by \pkg{IonSim.jl} package for simulating physical evolution of ion-based quantum computers. The package is built on the top of \pkg{QuantumOptics.jl}~\cite{kraemer2018quantumoptics.jl} numerical framework providing the functionality similar to \pkg{QuTiP}. However, \pkg{IonSim.jl} provides a specialized API for dealing with the dynamics of a configuration of trapped ions interacting with laser light. As such, it enables the emulation layer and can be used to develop and test quantum procedures targeting a specific hardware architecture. For example, given a specification of ion configuration and laser pulses, it is possible to utilize it as a virtual machine for higher-level layer, including quantum toolkits developed by hardware vendors or quantum programming languages.

\subsection{Quantum domain-specific languages}

The main trend observed recently in the field of quantum software is the development of software libraries enabling the embedding of elements used to create quantum code into a general purpose programming language. This approach has been taken by the companies developing dedicated hardware \cite{ibmqroadmap,xanadu-pennylane,rigetti-forest-sdk} as it allows them to control both hardware and software stack. Many of these software frameworks are based on \pkg{Python} programming language, which is especially convenient as quantum computing technologies are frequently related to applications in data science, an area where Python is frequently a first choice for developers. Thus, as quantum computing is frequently seen as a promising technology to boost data science, such frameworks naturally fulfill the needs of the community. 

However, the price to pay for the integration with \pkg{Python} libraries is twofold. Firstly, quantum SDKs usually have very limited capabilities concerning the simulation of quantum circuits, and one often needs to resort to dedicate software to execute computationally intensive part of software development. Secondly, \pkg{Python} has somehow limited capabilities concerning the symbolic computation. This factor is not limiting if one relies in their pipeline on the results obtained from quantum computers. However, it can be limiting if one needs to deal with the complicated optimization landscape and  aims at fine-tuning the ansatz used in a procedure like Variational Quantum Eigensolver or Quantum Approximate Optimization Algorithm.

\pkg{ProjectQ}~\cite{steiger2018projectq,haener2018software} delivers a software architecture for compiling quantum programs from a high-level language program to hardware-specific instructions. To some extent, this leverages the concept of specialized high-level languages, similar to \pkg{LanQ}~\cite{lanq} and \pkg{QCL}~\cite{oemerMSC1,oemerMSC2,oemerPHD} discussed below. However, the implementation is delivered as a \pkg{Python} library, which facilitates the usage of the software. The most important feature of \pkg{ProjectQ} is that it  enables the compilation of quantum algorithms and can be targeted at any specific quantum hardware implementation. Thus, it provides a hardware-agnostic solution. Unfortunately, it also presents a problem because such the project cannot expect to receive a dedicated support from any quantum hardware vendors.

\subsection{Quantum programming languages}

Quantum algorithms and communication protocols are developed in a~language of quantum circuits, where reversible logical operations are unitary matrices. While this method is convenient in the case of simple algorithms, it is very hard to operate on compound data types or if the algorithms are utilizing classical subroutines, which is a case for NISQ algorithms. This lack of abstraction motivated the development in the field of quantum programming languages. Until now, many quantum programming languages have been proposed, and it is natural to group them into two groups, reflecting two fundamentally different approaches -- imperative and functional -- in quantum software engineering.

\subsubsection{Imperative quantum programming}
The first family of quantum programming languages consists of the languages based on the imperative paradigm. Usually, these languages provide the syntax known from programming languages like \pkg{Pascal} and \pkg{C}. To support quantum computing, they extend standard syntax with the elements necessary to operate on quantum memory. 

In this group, \pkg{QCL} (Quantum Computation Language) \cite{oemerMSC1, oemerMSC2, oemerPHD} provides an example of one of the most advanced quantum programming language with working interpreter. Its syntax and classical data type hierarchy are based on \pkg{C}. In the original implementation, the programs written in \pkg{QCL} can be executed using the available interpreter. The interpreter is built on a top of \pkg{libqc} simulation library written in \pkg{C++} and offers an excellent speed of execution.

The second important example of quantum programming language based on the imperative paradigm is \pkg{LanQ}~\cite{lanq}. This language was developed to address the problems arising from the lack of elements supporting quantum communication. Additionally, \pkg{LanQ} was the first quantum programming language with full operation semantics specified~\cite{mlnarik2008semantics}. This allows the formal reasoning about the correctness of programs and for the further development of the language. Semantics is also crucial for the optimization of the programs written in \pkg{LanQ}.

It is interesting to notice that both above mentioned examples of imperative programming languages were developed as research projects. Unfortunately, the idea of delivering a specialized programming language, designed with fault-tolerant quantum computing resources in mind, did not gain any momentum in the NISQ era. This is surprising, as many aspects of \pkg{QCL} and \pkg{LanQ}  -- \eg quantum conditions and quantum links -- could be beneficial during the process of quantum software development. However, this suggests that a wider adoption is crucial for the success of quantum software framework. This, however, can only be achieved via harnessing a popularity of some traditional programming languages.

\subsubsection{Functional quantum programming}

Functional programming is often seen as an attractive alternative to traditional methods used in the field of computer techniques in science, which are mainly based on the imperative programming paradigm~\cite{hinsen2009promises}. Among the properties of functional programming that make it a more suitable approach for scientific computing and, in particular, for modelling quantum computing systems, are the transparency of the code produced in functional languages and the ease of code execution in parallel environments.

The potential of using functional programming languages for scientific computing and, in particular, for modelling quantum mechanical objects was first discussed in~\cite{karczmarczuk1999scientific}. Among the first attempts to introduce a functional programming style into the quantum domain, one can point to the extension of the lambda calculus~\cite{maymin1996extending}, quantum lambda calculus defined in~\cite{tonder2004lambda} and the development of an abstract environment based on a functional programming language to represent structures used in quantum mechanics~\cite{karczmarczuk2003structure,sabry2003modeling}. A more pragmatic approach is presented in~\cite{nyman2009symbolic}, where symbolic simulations of quantum algorithms in the \MMA environment are considered. However, similarly to the work on modelling quantum mechanical systems~\cite{karczmarczuk1999scientific,karczmarczuk2011specific}, the quantum programming languages based on the functional paradigm~\cite{miszczak2012high} developed so far do not support the key structures for modelling quantum information systems such as quantum channels. Thus, they have a very limited usability in the context of incorporating and mitigating quantum errors in the process of quantum computing. The existing tools and languages based on the functional paradigm have similar drawbacks, limiting their usability. This element is especially important in the NISQ era, where quantum machine is treated \textit{de facto} as a numerical subroutine, fine-tuned to cover only a very specific element of the computation.

Among the proposed functional quantum programming languages, there are many examples of experimental projects which focused on the theoretical aspects of quantum computational process. In particular, Quipper, proposed in \cite{green2013quipper,valiron2015programming}, was designed as  a scalable, expressive, functional, higher-order quantum programming language. More recently, QPCF~\cite{zorzi2019quantum,paolini2019qpcf} was proposed, with an approach to quantum programming based on functional calculi. In both of these cases, the actual implementation of the language is missing on incomplete. This severely limits the impact of the proposed approach on the current ecosystem of quantum software. Additionally, these approaches were mostly focused on the standard concept of quantum computation, based on the Quantum Random Access Machine (QRAM) model, which makes their adoption for NISQ quantum algorithms cumbersome. The efforts towards delivering quantum advantage using this type of tool are still ongoing. The architecture of the environment presented in the following part of this paper is to some degree motivated by the advances in functional quantum programming.

\subsection{Zero-code quantum programming}

One of the emerging options for creating quantum software is based on the recent trend of zero-code development. This type of quantum software is represented by the recently introduced \pkg{Pulsar Studio} ~\cite{pulsar-studio,pasqal-pulsar-s-1} solution, developed by French quantum hardware vendor Pasqal. \pkg{Pulsar Studio} enables the manipulation of Rydberg atoms, the technology underlying the hardware developed by the company. As in the case of \pkg{IonSim.jl}, the provided API is fine-tuned to the capabilities offered by the underlying physical phenomena. Additionally, as in that case, the code is targeting a specific hardware, \pkg{Pular Studio} imposes limitation resulting from the physical characteristics of the hardware platform. On the other hand, the programmer does not need to deal with the internal details of the physical evolution.

\section{Environment for symbolic quantum programming}
\label{sec:towards}

Following the review of the quantum software ecosystem, we now focus on the elements required to deliver a software environment addressing the some of the outlined limitation of the existing toolkits and frameworks. To this end, we start by listing the requirements for such an environment, reflecting the experience with developing such tools. We focus on the symbolic capabilities, which are important for harnessing the theoretical aspects of quantum theory to develop robust quantum programs. We also describe the desired architecture of the environment, especially considering the hybrid character of quantum variational algorithms. 

\subsection{Requirements for quantum programming toolkits}

In particular, the environment supporting symbolic development of quantum programs should address the following issues:
\begin{itemize}
	
	\item \textbf{Stability and long-term support.} This element is commonly missing in the existing software packages and toolkits. In some cases, this is due to the dynamics of progress of the software and hardware development. Nevertheless, the lack of stability is a serious hurdle for reproducing research results and for developing the elements of the quantum software ecosystem such as external libraries of procedures.
	
	\item \textbf{Support for symbolic computing.} Such support is required to provide capabilities to deal with the structure of the space of quantum states and quantum channels. It is also needed for providing support for methods based on information geometry, commonly used in hybrid quantum-classical algorithms to deal with the optimization of the energy landscape.
	
	\item \textbf{Versatility of general purpose programming language.} As it was pointed in \secref{sec:stat-of-art}, a plethora of existing solution is based on general purpose languages. For the purpose of this work, we advocate the usage of functional paradigm. Fortunately, during the last few years the elements of functional programming have become more and more popular, and this programming paradigm is now supported by many mainstream programming languages. 
	
	\item \textbf{Extensibility capabilities.} There are several justifying this requirement. Such capabilities are essential for enabling the integration with high-performance numerical libraries and communication with external services, including cloud-based resource providers~\cite{awsquantum} or other resources, \eg sources of randomness~\cite{miszczak2013employing,miszczak2013randfile}. Additionally, they are also crucial for enabling the access to external resources, including access to quantum computer emulators like \pkg{IonSim.jl}, which are -- from the perspective of computing pipeline -- treated as specialized numerical coprocessors. 
	
	\item \textbf{Support for high-efficiency numerical procedures.} This element is important to provide the means for conducting numerical simulations of small- and middle-size quantum programs. In many cases, this element is supported thanks to the library of functions or directly provided by the hosting computer algebra system. One should note that the ability to run such simulation is essential for some software engineering applications, including quantum software testing.
	
	\item \textbf{Support for parallel and distributed computing.} Such support provides some advantages concerning the possibility of simulating quantum systems. This is especially important considering the approach similar to quantum knitting, proposed by IBM~\cite{ibm-circuit-knitting,piveteau2022circuit}, which implies the possibility of distributing the computation required to simulate the next generation of quantum computers.
	
\end{itemize}

\subsection{Implementation of symbolic quantum programming}

Let us now present a structure of a quantum software framework supporting the functional programming and symbolic manipulation of quantum objects~\cite{qi-repo}. The framework has been developed in the form of a package written in Wolfram language, running in \MMA computer algebra system. In the initial phase, the package was developed to support the symbolic analysis of quantum states and operations only~\cite{miszczak2011singular,miszczak2012generating}. 

Among the structures required by any quantum software framework one 
can distinguish three major categories of basic structures focused on the following aspects:

\begin{itemize}
	\item \textit{Logic and computation} -- constructions for managing quantum memory and processing coherent quantum information;
	\item \textit{Physics} -- elements for dealing with decoherence and quantifying quantum mechanical properties of the systems;
	\item \textit{Extendability and connectivity} -- functionality enabling communication with external quantum resources and providing building blocks for hybrid quantum-classical algorithms.
\end{itemize}

We also include three main categories of the constructions used in the framework: \textit{basic structures} dealing with elementary components used in quantum computing, \textit{computational structures} providing support for high-level programming construction and physical concepts, and \textit{variational structures} for dealing with hybrid quantum-classical model of computation commonly used in variational quantum algorithms.

Among the functions described below, most of the \textit{basic structures} and \textit{computational structures} are included in the \pkg{QI} package. Selected \textit{variational structures} are implemented in \pkg{QINisq} packages. Additionally, some additional functionality is provided in \pkg{QIExtras} package. A detailed description is included in the documentation available from the project repository~\cite{qi-repo}.

\subsubsection{Logic and computation}

\paragraph{Basic structures} On the most elementary level each quantum software framework has to deal with the local manipulation of qubits, \ie the elementary units of quantum information, and to include the elements required to allocate quantum memory and to manipulate the state of the qubits. Among these elements, clearly, one can include \mma{Ket}, \mma{Bra} and \mma{Proj} functions, supporting the qubit-level manipulation of quantum memory. However, the support for symbolic quantum programming requires also the implementation of the  functions for manipulating a general form of quantum mixed states and state vectors, including \mma{SymbolicMatrix} and \mma{SymbolicVector}, as well as functions for defining quantum gates, \eg \mma{SpecialUnitary}. Additionally,  specialized functions exploiting the theoretical structure of quantum objects in the process of program development, such as \mma{SymbolicHermitianMatrix} and \mma{SymbolicBistochasticMatrix} are required. Utilization of such specialized types is beneficial for testing the developed programmes -- for example, by ensuring that the evolution preserves some of the properties of the input state. Moreover, by clearly defining the domain of the processed object, one can support the inference of analytical results concerning the processed objects.

\paragraph{Computational structures} Building upon the basic structures, including qubits and unitary gates, the logical structures supporting computation are represented by functionality for managing registers, building complex quantum gates, and managing controlled quantum operators. Such functions include \mma{KroneckerProduct}, \mma{Gate}, and \mma{CGate}. This group also includes the functionality providing general templates for common quantum algorithms, including standard quantum algorithms like Quantum Fourier Transform, as well as variational quantum algorithms like Quantum Approximate Optimization Algorithm.

\paragraph{Variational structures} This group of functions is responsible for handling measurement results, managing objective functions, and for the interaction with classical optimizer in the variational loop. In this case, the utilization of symbolic computation can be applied to boost the quality of the optimization results. Moreover, thanks to the symbolic capabilities, the structure of the optimization landscape can be utilized to optimize the structure of the ansatz used in the quantum variational algorithms. 

\subsubsection{Physics}

\paragraph{Basic structures}  On the basic level, the functionality related to the physics behind quantum computing is related with the structure of the dynamics of quantum systems. Functions in this category include functions for managing Kraus representation, including \mma{SuperoperatorToKraus}, \mma{Superoperator}, and \mma{ApplyChannel}. These functions can be used to provide an error model and incorporate it into the computation process.

\paragraph{Computational structures} Functionality in this area is focused on the composite system used during the process of quantum computation. An example of function supporting this process is \mma{ProductSuperoperator}, which can be used to apply noise model using the structures of quantum register as opposite to singe qubits. This group of functions also includes the functions for handling permutations of quantum registers and classically controlled quantum operations. In this case, the functional programming enables very clear and concise implementation of operations such as \mma{PartialTrace} or any other operations implemened on a subset of qubits or quantum sub-registers. More details can be found in~\cite{miszczak2011singular}.  

\paragraph{Variational structures} The physical aspects of quantum hardware can be included by managing the specification provided by hardware vendors. Alternatively, one can assume some simplified error model and utilize the computational structures supporting physics mentioned above to build the error model mimicking the realistic scenario. Another option for managing these aspects of computation is to utilize randomizer processing. This can  be achieved by utilizing by using functions providing interface for managing quantum random objects, including \mma{RandomKet}, \mma{RandomUnitary}, and \mma{RandomDynamicalMatrix}. This type of approach was utilized for developing the package described in~\cite{miszczak2012generating}. 

\subsubsection{Extendability and connectivity}

The last functional group expected to be delivered by any quantum software framework requires support for extendability in the host programming language or computer algebra system. There are many examples of such modules, including \pkg{qiskit.providers} API developed by IBM~\cite{qiskit-providers-docs}. In the case of \MMA computer algebra system, quantum modules harnessing the access to external resources were described in~\cite{miszczak2012generating, miszczak2013employing}. 

\paragraph{Basic structures} On the basic level, the functionality in this group includes, for example, the management of external sources of  randomness. This type of functionality is obligatory, as most of the features delivered by these functions can be emulated using the capabilities of the host. In the context of the described packages, the functionality for managing remote sources of randomness for the purpose of utilizing it in the quantum stat analysis was described in \cite{miszczak2013employing}. Another example is the utilization of off-line true random sources. In this case, some attention has to be devoted to managing random numbers without compromising the quality of the generated random objects~\cite{miszczak2013randfile}.

\paragraph{Computational structures} This functionality covers the function implementing non-physical transformations (\ie maps that cannot be realized in laboratory) of quantum objects. The best known examples are \mma{Resuffling} and \mma{PartialTransposition}. Another example is \mma{TruncatedFidelity}, which can be used as an objective function in variational quantum computation and is based on the measurement results from a quantum device.

\paragraph{Variational structures} Procedures in this group are used for the management of quantum computers, including the definitions and specification of backbends used to execute the code, and the management of quantum memory using classical control structures. This group of functions delivers means for managing quantum execution, including \mma{RunGate}, \mma{RunCGate}, and \mma{QRun}. It should also provide means for managing backbends, analysing the results, providing feedback to the classical optimization subroutines, and possibly modify the structure of the ansatz. For the case of classical optimization and ansatz optimization, the external subroutines are usually employed to handle the requested functionality. In such case, the additional information about the current structure could be used to improve the operations of the external subroutine.

\section{Applications for quantum computing technologies}
\label{sec:applications}

The architecture and preliminary implementation provided in the previous section addresses some of the issues one can encounter as a quantum software user. Accordingly, some of the remarks about the advantages and disadvantages of the existing approaches are based on the author's experience with implementing and running quantum code. However, to provide some wider perspective on this subject, we include two used cases where quantum software is of the central importance for the further progress. The first problem is related to the process of finding the best ansatz for the given problem or class of problems. The second problem is focused on the construction of test cases and quantum software testing.

\subsection{Circuit construction and optimization}

One of the major problems in the process of developing quantum algorithms which could achieve quantum advantage are the limitations of the currently available quantum hardware -- NISQ computers. This type of machine is limited in terms of qubit connectivity and coherence time. Thus, any quantum variational algorithm targeting such machine has to be optimized by considering these limitations. In most cases, the methods employed for this process require a human expert possessing knowledge about the structure of individual qubits and the interactions between them. However, as the size of the developed quantum computers is growing rapidly, this task is becoming unattainable. 

For this reason, the methods employing machine learning approach have been recently proposed. One of the proposed approaches includes the  utilization of reinforcement learning techniques. In particular, in~\cite{ostaszewski2021reinforcement} it was demonstrated that Deep Reinforcement Learning (DRL) approach can successfully solve the quantum circuit construction problem. 

However, the existing reinforcement learning methods are limited to very specialized cases, usually involving a fixed problem to execute on a particular machine or a fixed noise model. Due to the high computational overhead, most studies do not consider a scenario with changing environment, including device noise. Hence, the proposed methods require the repetition of the learning process as soon as the characteristic of the system-environment interactions changes. Thus, it would be desirable for the machine learning methods to continually adapt to the changing noise patterns. For this, it is necessary to provide data about the impact of the changes in the obtained outcomes as a dependency on the changes in the ansatz structure. This usually requires a significant number of test runs. 

The utilization of symbolic computing for this class of problems is very similar to the case of exploring the optimization landscape in the case of variational quantum computing. Thus, by supporting this information with the symbolic computation, one can possibly speed up the process of exploring the space of possible solutions. What's more, the ability to reproduce a model of a quantum machine and obtain an analytical form of the objective could be used directly to emulate the process of quantum computation, significantly decreasing the computational power requirements.

\subsection{Quantum Software testing}

The problem of testing the correctness of quantum software provides another usage example where the symbolic evaluation of quantum computing could provide important benefits. As in the case of standard software, testing quantum software aims at delivering automated  and systematic methods to guarantee the correctness of the  quantum code. Such a guarantee is critical in the case of quantum software for two main reasons. 

Firstly, in the case of quantum computing, the internal structure of memory cannot be observed and, at the same time, it is more complicated due to the significantly larger space of available states in comparison to the classical computation. Hence, the correctness of the programs has to be inferred from the limited information about the process of computation. This is especially problematic because for the large class of internal states, occurring in the algorithms related to modelling quantum chemistry phenomena, the state of quantum processor cannot be efficiently represented and analysed. 

Secondly, even for the relatively modest size quantum computers, the process of simulating the quantum information processing is computationally demanding. This problem is related to the issue of the internal stat representation. It also implies that the process of testing of middle size quantum computers cannot be easily delegated to the simulator. Hence, the methods for generalizing the testing obtained for the quantum subsystem of quantum computers -- partial quantum testing -- could be used to address this issue. 

Some solutions proposed recently address the problem of quantum software testing. In \cite{nan2022quantum} a method of quantum symbolic was proposed to generate test cases, which helps finding bugs in quantum programs. In \cite{mendiluze2021muskit}, a quantum mutation analysis tool \pkg{Muskit} for  domain-specific language defined in \pkg{Qiskit} was introduced. Another tool is \pkg{Quito}~\cite{wang2021quito}, also targeting \pkg{Qiskit}, and enabling the automatic generation of test suites, covering three coverage criteria defined on inputs and outputs. Both \pkg{Muskit} and \pkg{Quito} require from the user the knowledge about the input-output characteristics of quantum device, which is not always feasible.

By utilizing the symbolic representation of quantum programs, one can easily extend the usability of the existing quantum software approaches. For example, by utilizing the symbolic form of the test cases, it is possible to extend the range of generated test cases and include non-classical state vectors with specific properties. Additionally, by inferring the form of the partial evolution, it is possible to generate test cases targeting only the selected quantum subsystems. This could, to some degrees, overcome the problem of testing the full space of quantum states and strengthen the validation results.

\section{Conclusions}
\label{sec:conclusions}

The goal of the presented paper was to provide an overview on the quantum software development from the perspective of quantum technologies and to describe some crucial elements one has to support to fully harness the potential of quantum computing via quantum software. One of the main observations one should take home after reading this manuscript is that the quantum software ecosystem is very fragmented, and many ideas for delivering quantum software are considered and implemented. This is understandable, as the quantum computing is still a very new field of computing. For the user, however, this introduces difficulties related with code stability, documentation, and community support.

On the other hand, from the perspective of software engineering, some of the most important issues with quantum code are hard to grasp -- those include the structure of quantum states and the non-classical constructions used to develop quantum code. Hence, we would like to emphasize the importance of software engineering education for technologies based on quantum computing. And to support education, stable and reliable quantum software is needed.

The applications of quantum mechanics to provide computational advantage were among the first studied by theoretical physicists and computer scientists. Currently, however, one cannot simply claim that the proposed algorithm or protocol can achieve theoretical advantage and for advancing the state of quantum computing technology there is a strong requirement to deliver algorithms which can achieve quantum advantage and beat classical algorithms in realistic use cases. This, however, requires quantum algorithms which can be run on realistic hardware, usually significantly limited in terms of memory and accuracy. The quantum code executed on such hardware has to be fine-tuned to meet its limitations, and tested to check its accuracy. None of these task cannot be done manually and require the software supporting quantum programmers.

One of the important aspects of quantum computing technologies is related to the interplay between the power of quantum computing in realistic scenarios, expressed in possible advantages of utilizing quantum machines, and the impact of intrinsic inaccuracy present in quantum computers. This inaccuracy is commonly described as quantum noise, but it can also be related to the limitations of quantum computers, as in most case the requested logical operations cannot be executed with arbitrary precision. This leads to the unavoidable error, which will be always present in the executed quantum programs. For this reason, it is crucial to understand to what degree this intrinsic inaccuracy has impact on the results obtained using quantum computers, and to provide methods for counteracting it. Quantum symbolic programming enables the programmer to gain in-depth insight into the structure of the quantum program. Such insight is crucial to fully exploit the structure of quantum theory for achieving quantum advantage.

\paragraph{Acknowledgements}
The author would like to acknowledge valuable comments on the problem of quantum architecture search from Mateusz Ostaszewski and Akash Kundu, and motivating discussions with Adam Glos, \"Ozlem Salehi, and Ludmila Botelho concerning quantum software ecosystem. This work received support from the Polish National Science Center under the grant agreement 2019/33/B/ST6/02011 and from Polish National Information Processing Institute  under \emph{National Supercomputing Infrastructure for EuroHPC -- EuroHPC PL} project funded at the Smart Growth Operational Programme (2014-2020), grant agreement POIR.04.02.00-00-D014/20-00.



\end{document}